\documentclass[12pt]{article}
\usepackage{eurosym}
\usepackage{graphicx}
\usepackage{pdfpages}
\usepackage{authblk}
\usepackage[numbers,sort&compress]{natbib}
\usepackage{xcolor}
\usepackage[utf8]{inputenc}
\usepackage{amsmath}
\usepackage{geometry}
\usepackage{amsmath, amssymb, geometry}
\usepackage{float}
\usepackage{tikz}
\usepackage{graphicx}
\usepackage{braket}
\title{Multi-Soliton Propagation and Interaction in $\Lambda$-Type EIT
Media: An Integrable Approach}

\author[1]{Ramesh Kumar Vaduganathan$^{\dagger}$}
\author[2]{Prasanta K. Panigrahi$^\ddagger$}
\author[3]{Boris A. Malomed$^*$}

\affil[1]{Department of Physics, Velammal Engineering College (Autonomous), Chennai-600066, India}
\affil[2] {Center for Quantum science and Technology (CQST), Siksha ‘o’ Anusandhan university, Bhubaneswar, 751030, India} 
\affil[2]{Department of physical sciences, IISER-Kolkata, Mohanpur, Nadia, 741246, India}

\affil[3]{Department of Physical Electronics, School of Electrical Engineering, Faculty of Engineering, and Center for Light-Matter Interaction, Tel Aviv P.O. Box 39040, Israel}
\affil[3]{Instituto de Alta Investigaci\'{o}n, Universidad de Tarapac\'{a}, Casilla 7D, Arica, Chile}

\begin{document}
\maketitle

\begin{center}
Email: ramehkumar@velammal.edu.in$^{\dagger}$, pprasanta@iiserkol.ac.in $^{\ddagger}$, malomed@tauex.tau.ac.il$^{*}$
\end{center}

\begin{abstract}
Electromagnetically induced transparency (EIT) is well known as a quantum
optical phenomenon that permits a normally opaque medium to become
transparent due to the quantum interference between transition pathways.
This work addresses multi-soliton dynamics in an EIT system modeled by the
integrable Maxwell-Bloch (MB) equations for a three-level $\Lambda $-type
atomic configuration. By employing a generalized gauge transformation, we
systematically construct explicit N-soliton solutions from the corresponding
Lax pair. Explicit forms of one-, two-, three-, and four-soliton solutions
are derived and analyzed. The resulting pulse structures reveal various
nonlinear phenomena, such as temporal asymmetry, energy trapping, and
soliton interactions. They also highlight coherent propagation, elastic
collisions, and partial storage of pulses, which have potential implications
for the design of quantum memory, slow light and photonic data transport in
EIT media. In addition, the conservation of fundamental physical quantities,
such as the excitation norm and Hamiltonian, is used to provide direct
evidence of the integrability and stability of the constructed soliton
solutions.
\end{abstract}

\section{Introduction}

A weak probe laser can penetrate an initially opaque medium when a strong
coherent coupling field is present, a phenomenon known as the
electromagnetically induced transparency (EIT) \cite%
{Harris1990,Boller1991,Harris1997,Fleischhauer2005}. In multi-level atomic
systems, this phenomenon originates from destructive interference between
excitation channels. It has attracted a lot of interest because of its
crucial role in regulating light-matter interactions. Slow light, optical
memory, nonlinear switching, and quantum data processing are optical
phenomena and applications that are supported by EIT \cite%
{Hau1999,Kasapi1995, Liu2001, Phillips2001, Bajcsy2003}.

One of the most important models for understanding EIT is the three-level $%
\Lambda $-type atomic configuration, in which the combined evolution of the
atomic population, coherence, and electromagnetic fields is governed by the
system of Maxwell-Bloch (MB) equations. Under the conditions of the slowly
evolving envelope and rotating-wave assumption, these equations characterize
the coupled evolution of the optical fields and atomic coherence \cite%
{Maimistov1984,Park1998,Wadati2008,Kumar2008,lakshmanan2015}. The MB system becomes
integrable under certain resonance conditions, enabling the existence of
soliton solutions.

The seminal work \cite{Wadati2008} has shown that simulton-type soliton
solutions are produced by the integrable MB system. Depending on the initial
population distribution between the atomic states, the system exhibits many
nonlinear phenomena, such as energy sharing, optical switching, transparency
windows, and the pulse storage provided by adiabatically turning off the
coupling beam \cite{Lukin2003, Wu2010, Ling1998, Duan2001,
Chaneliere2005,Hedges2010,Zhao2009,Sangouard2011,Simon2010,Hammerer2010,Ma2017, Novikova2012}%
. The feasibility of using the relative phase and incoherent pumping to
shape the optical response in quantum media has been demonstrated by recent
studies, that have also examined the role of spontaneously generated
coherence and phase control in achieving the all-optical switching and
coherent pulse manipulation in $\Lambda $-type systems \cite{Hien2022}.

In spite of these advancements, there is still a lack of systematic analysis
of multi-soliton solutions in the MB framework, particularly as concerns
multi-pulse EIT dynamics that are essential for developing quantum network
architectures. In this work, we aim to partly fill this gap by producing
explicit N-soliton solutions of the MB system in the $\Lambda $-type EIT
medium by means of a generalized iterative gauge transformation \cite%
{Chau1991}. This method suggests a possibility to design scalable
optical-data processing, by means of controlled creation of intricate
soliton structures and evolution patterns.

Although the present formulation is strictly integrable under ideal resonance, realistic photonic and EIT platforms generally operate under conditions where small detuning, dephasing, or loss are inevitable. The multi-soliton states derived here provide analytical benchmarks that can serve as initial conditions or validation references for numerical simulations in such non-integrable regimes. Therefore, the present integrable analysis offers a theoretical guideline for exploring robust pulse propagation and coherent control in realistic quantum-optical systems.

The Maxwell-Bloch equations for the EIT configuration are formulated in
section 2 of this study, along with the respective integrability conditions.
Using the generalized gauge transformation, we iteratively generate explicit
N-soliton solutions in section 3. In section 4 we build one-, two-, three-,
and four-soliton solutions by dint of this method, and in section 5 we
examine their spatiotemporal dynamics. In section 6 we highlight the
potential of multi-soliton interactions to enable improved photonic
functions in EIT-based systems, and address the ramifications of these
findings for the quantum data transfer and light storage, supported by the
conserved quantities. The paper is concluded by section 7.

\section{The MB\ (Maxwell--Bloch) equations}

The MB equations describe the interaction of light with a three-level $%
\Lambda $-type atomic system under the EIT\ condition. These equations
relate the density-matrix components, that represent the atomic-state
populations and coherence, to the evolution of the probe's and
coupling-beams' electric-field envelopes.

We consider the $\Lambda $-type atomic configuration including three levels,
\textit{viz}., the ground state $\ket{1}$, metastable state $\ket{2}$, and
the excited state $\ket{3}$. A weak probe field with electromagnetic
frequency $\omega _{1}$ is tuned to the $\ket{1}\leftrightarrow \ket{3}$
transition, and a strong coupling field of frequency $\omega _{2}$ drives
the $\ket{2}\leftrightarrow \ket{3}$ transition. Under the slowly varying
envelope approximation, and assuming that the pulse durations are much
shorter than the atomic relaxation times, the MB system is written as \cite%
{Maimistov1984,Park1998,Wadati2008,Kumar2008}
\begin{align}
\frac{\partial \mathcal{E}_{1}}{\partial z}+\frac{n}{c}\frac{\partial
\mathcal{E}_{1}}{\partial t}& =-i\frac{2\pi N\hbar \omega _{1}}{cn}\mu
_{13}\rho _{13},  \label{eq:MB1} \\
\frac{\partial \mathcal{E}_{2}}{\partial z}+\frac{n}{c}\frac{\partial
\mathcal{E}_{2}}{\partial t}& =-i\frac{2\pi N\hbar \omega _{2}}{cn}\mu
_{23}\rho _{23},  \label{eq:MB2} \\
i\hbar \frac{\partial \rho }{\partial t}& =[H,\rho ],  \label{eq:MB3}
\end{align}%
where $\mathcal{E}_{1}$ and $\mathcal{E}_{2}$ are the complex envelopes of
the probe and coupling fields, respectively. Further, $N$ is the atomic
density, $n$ is the refractive index, $\mu _{ij}$ = $-d_{ij}/\hbar $ are
elements of the electric-dipole matrix for the atomic transitions, and $\rho
$ is the $3\times 3$ density matrix of the atomic system.

The Hamiltonian that governs the dipole atomic dynamics under the
rotating-wave approximation is
\begin{equation}
H=\hbar
\begin{pmatrix}
0 & 0 & \mu _{13}\mathcal{E}_{1}^{\ast } \\
0 & \Delta \omega _{1}-\Delta \omega _{2} & \mu _{32}\mathcal{E}_{2}^{\ast }
\\
\mu _{31}\mathcal{E}_{1} & \mu _{32}\mathcal{E}_{2} & \Delta \omega _{1}%
\end{pmatrix}%
,
\end{equation}%
where $\Delta \omega _{1}=\omega _{31}-\omega _{1}$ and $\Delta \omega
_{2}=\omega _{32}-\omega _{2}$ are the Rabi frequencies of the probe and
coupling fields, and $\Delta $ is the detuning parameter.

To simplify the system, we introduce the traveling coordinate, $x=t-nz/c$,
and the rescaled propagation variable, $T=z/l$, which allows cast the system
in a more symmetric form. With suitable rescaling of the fields, $q_{1}=\mu
_{31}\mathcal{E}_{1}$ and $q_{2}=\mu _{32}\mathcal{E}_{2}$, and adopting the
resonance condition ($\Delta =0$), the system becomes
\begin{align}
\frac{\partial q_{1}}{\partial T}& =-i\rho _{31}, \\
\frac{\partial q_{2}}{\partial T}& =-i\rho _{32}, \\
i\hbar \frac{\partial \rho }{\partial x}& =[H_{\text{res}},\rho ],
\end{align}%
where $H_{\text{res}}$ is the Hamiltonian corresponding to the resonance
condition
\begin{equation}
H_{\text{res}}=\hbar
\begin{pmatrix}
0 & 0 & q_{1}^{\ast } \\
0 & 0 & q_{2}^{\ast } \\
q_{1} & q_{2} & 0%
\end{pmatrix}%
.
\end{equation}

This system becomes integrable under specific conditions and admits the
Lax-pair representation, enabling one to use analytical techniques, such as
the inverse scattering transform \cite{Ablowitz1991}, B\"{a}cklund transform
\cite{Miura1978}, and gauge/Darboux transform \cite{Matveev1991,Gu2006}, to
produce soliton solutions \cite{Drazin1989}. In this paper, we apply the
gauge-transformation approach to systematically construct higher-order
soliton solutions, which reveal diverse dynamical effects.

\section{Constructing N-soliton solutions by means of the gauge transform}

To obtain soliton solutions of the MB equations governing the EIT system, we
utilize the integrability of the reduced system under the resonance
condition and apply a generalized gauge transform \cite{Chau1991}. This
method makes it possible to systematically generate exact N-soliton
solutions by successively applying elementary transformations to the vacuum
solution of the associated Lax pair.

\subsection{Lax-pair operators}

The integrable MB system, based on Eqs.(5)-(8), admits the following
Lax-pair representation,
\begin{align}
\frac{\partial \Psi }{\partial x}& =U(x,T;\lambda )\Psi ,  \label{eq:laxU} \\
\frac{\partial \Psi }{\partial T}& =V(x,T;\lambda )\Psi ,  \label{eq:laxV}
\end{align}%
where $\lambda $ is a complex spectral parameter, with the operators
\begin{align}
U(x,T;\lambda )& =%
\begin{pmatrix}
i\lambda & 0 & -iq_{1}^{\ast } \\
0 & i\lambda & -iq_{2}^{\ast } \\
-iq_{1} & -iq_{2} & -i\lambda%
\end{pmatrix}%
, \\
V(x,T;\lambda )& =\frac{i}{2\lambda }%
\begin{pmatrix}
\rho _{11} & \rho _{12} & \rho _{13} \\
\rho _{21} & \rho _{22} & \rho _{23} \\
\rho _{31} & \rho _{32} & \rho _{33}%
\end{pmatrix}%
.
\end{align}%
The compatibility condition, $\partial _{T}U-\partial _{x}V+[U,V]=0$,
recovers the integrable system of equations (5)-(7) that govern the
evolution of the fields and atomic population in the EIT system.

\subsection{The trivial seed solution}

To generate soliton solutions by means of the gauge transformation, we begin
with the trivial (vacuum) seed solution. It corresponds to the situation in
which the probe and coupling fields are initially absent:
\begin{equation*}
q_{1}(x,T)=0,\quad q_{2}(x,T)=0.
\end{equation*}%
Additionally, the atomic system is assumed to be in the incoherent state,
with all off-diagonal density matrix elements vanishing,
\begin{equation}
\rho _{jk}(x,T)=0,\quad j\neq k,
\end{equation}%
solely the populations in each state being nonzero:
\begin{equation}
\rho _{jj}(x,T)=\sigma _{jj},\quad j=1,2,3.
\end{equation}%
This initial condition defines the vacuum background of the system, which is
employed, as usual, as the seed for constructing soliton solutions. In this
case, the Lax pair becomes a diagonal system:
\begin{align}
U^{[0]}& =i\lambda J=%
\begin{pmatrix}
i\lambda & 0 & 0 \\
0 & i\lambda & 0 \\
0 & 0 & -i\lambda%
\end{pmatrix}%
, \\
V^{[0]}& =\frac{i}{2\lambda }%
\begin{pmatrix}
\sigma _{11} & 0 & 0 \\
0 & \sigma _{22} & 0 \\
0 & 0 & \sigma _{33}%
\end{pmatrix}%
.
\end{align}%
The eigenfunction of the Lax pair under this vacuum condition is
\begin{equation}
\Psi ^{\lbrack 0]}(x,T;\lambda )=%
\begin{pmatrix}
e^{i\lambda x+\frac{i}{2\lambda }\sigma _{11}T} & 0 & 0 \\
0 & e^{i\lambda x+\frac{i}{2\lambda }\sigma _{22}T} & 0 \\
0 & 0 & e^{-i\lambda x+\frac{i}{2\lambda }\sigma _{33}T}%
\end{pmatrix}%
.
\end{equation}%
This simple solution provides an appropriate starting point for constructing
soliton solutions through successive iterations of the gauge transformation.

\subsection{The N-fold gauge transformation}

To construct N-soliton solutions, we apply an N-fold gauge transformation,
\begin{equation}
\Psi ^{\lbrack N]}(x,T;\lambda )=G_{N}(\lambda )\Psi ^{\lbrack
0]}(x,T;\lambda ),
\end{equation}%
where the gauge matrix $G_{N}(\lambda )$ is built iteratively from the
rank-one projection matrix (i.e., $P^{2}=P$):
\begin{equation}
G_{N}(\lambda )=\prod_{j=1}^{N}\left( I-\frac{\lambda _{j}-\lambda
_{j}^{\ast }}{\lambda -\lambda _{j}^{\ast }}P_{j}\right) ,  \label{eq:gauge}
\end{equation}%
with
\begin{equation}
P_{j}=\frac{\Psi ^{\lbrack j-1]}(\lambda _{j})\otimes \Psi ^{\lbrack
j-1]\dagger }(\lambda _{j})}{\Psi ^{\lbrack j-1]\dagger }(\lambda _{j})\Psi
^{\lbrack j-1]}(\lambda _{j})},
\end{equation}%
and $\Phi _{j}=\Psi ^{\lbrack j-1]}(x,T;\lambda _{j})\cdot \mathbf{v}_{j}$,
where $\mathbf{v}_{j}$ is an arbitrary constant polarization vector, $%
\mathbf{v}_{j}=(\varepsilon _{1j},\varepsilon _{2j},1)^{\mathrm{tr}}$. and $%
I $ is the identity matrix.

At each stage, the fields $q_{1}$ and $q_{2}$ are updated recursively:
\begin{align}
q_{1}^{[j]}& =q_{1}^{[j-1]}-2i(\lambda _{j}-\lambda _{j}^{\ast })\frac{%
(P_{j})_{31}}{\Phi _{j}^{\dagger }\Phi _{j}}, \\
q_{2}^{[j]}& =q_{2}^{[j-1]}-2i(\lambda _{j}-\lambda _{j}^{\ast })\frac{%
(P_{j})_{32}}{\Phi _{j}^{\dagger }\Phi _{j}}.
\end{align}%
These recursion relations start from the vacuum state $%
q_{1}^{[0]}=q_{2}^{[0]}=0$, with the respective projector,
\begin{equation}
P_{j}=\frac{\Phi _{j}\Phi _{j}^{\dagger }}{\Phi _{j}^{\dagger }\Phi _{j}}.
\end{equation}

\subsection{The final form of the N-soliton solution}

After performing $N$ gauge transformations, the resulting soliton solution
for the probe and coupling fields can be expressed, in a compact form, as
\begin{align}
q_{1}^{[N]}(x,T)& =-\sum_{j=1}^{N}2i(\lambda _{j}-\lambda _{j}^{\ast })\frac{%
(\Phi _{j}\Phi _{j}^{\dagger })_{31}}{\Phi _{j}^{\dagger }\Phi _{j}},
\label{eq:q1N} \\
q_{2}^{[N]}(x,T)& =-\sum_{j=1}^{N}2i(\lambda _{j}-\lambda _{j}^{\ast })\frac{%
(\Phi _{j}\Phi _{j}^{\dagger })_{32}}{\Phi _{j}^{\dagger }\Phi _{j}}.
\label{eq:q2N}
\end{align}%
It clearly reveals the additive contribution of each individual soliton to
the total field and ensures the preservation of the integrability and
coherence properties at each step. The spectral parameters $\lambda _{j}$
and polarizations $\mathbf{v}_{j}$ control the velocity, amplitude, and
relative phase of each soliton component.

It is relevant to clarify the novelty of the results reported here. The one-
and two-soliton solutions of the MB system were discussed previously in
Refs. \cite{Wadati2008,Kumar2008}, where their explicit form and basic
interaction properties were analyzed. In contrast, to the best of our
knowledge, the exact three- and four-soliton solutions have not been
reported before. The present work therefore extends the known results by
constructing the closed-form expressions for higher-order soliton states ($%
N>2$) in the MB EIT system. These solutions are derived here systematically,
using the gauge-transformation method, and their physical properties,
including conserved quantities, energy distribution, and interaction
dynamics, are analyzed in detail. Thus, our results provide the
comprehensive characterization of multi-soliton dynamics ($N=3,4$) in this
integrable light-matter system.

\section{The construction of multi-soliton solutions}

\subsection{The one-soliton solution}

From the general N-soliton solution,
\begin{equation}
q_{k}^{[N]}(x,T)=-\sum_{j=1}^{N}2i(\lambda _{j}-\lambda _{j}^{\ast })\frac{%
(\Phi _{j}\Phi _{j}^{\dagger })_{3k}}{\Phi _{j}^{\dagger }\Phi _{j}},\quad
k=1,2,
\end{equation}%
we begin with the general one-fold (applying the gauge transform once, which
generates the one-soliton solution from the seed (vacuum) state)
gauge-transformed solution of the MB system:
\begin{equation*}
q_{k}^{[1]}(x,T)=-2i(\lambda _{1}-\lambda _{1}^{\ast })\frac{(\Phi _{1}\Phi
_{1}^{\dagger })_{3k}}{\Phi _{1}^{\dagger }\Phi _{1}},\quad k=1,2.
\end{equation*}%
We thus obtain the one-soliton solution ($N=1$) as
\begin{align}
q_{1}^{[1]}(x,T)& =-2i(\lambda _{1}-\lambda _{1}^{\ast })\frac{(\Phi
_{1}\Phi _{1}^{\dagger })_{31}}{\Phi _{1}^{\dagger }\Phi _{1}}, \\
q_{2}^{[1]}(x,T)& =-2i(\lambda _{1}-\lambda _{1}^{\ast })\frac{(\Phi
_{1}\Phi _{1}^{\dagger })_{32}}{\Phi _{1}^{\dagger }\Phi _{1}},
\end{align}%
where $\Phi _{1}=\Psi ^{\lbrack 0]}(x,T;\lambda _{1})\mathbf{v}_{1}$, and $%
\mathbf{v}_{1}=(\varepsilon _{11},\varepsilon _{21},1)^{T}$ is a constant
polarization vector. Using vacuum eigenfunction $\Psi ^{\lbrack 0]}$ given
by Eq. (18) and applying the gauge transformation, we obtain the one-soliton
solution in the following form:

\begin{align}
q_1^{[1]}(x,T) &= -\frac{4\sqrt{2}\, \beta_1 \varepsilon_{11} e^{\Theta_1}}{%
2e^{\Xi_1} + e^{\Xi_2} + e^{\Xi_3}}, \\
q_2^{[1]}(x,T) &= -\frac{4\sqrt{2}\, \beta_1 \varepsilon_{21} e^{\Theta_2}}{%
2e^{\Xi_1} + e^{\Xi_2} + e^{\Xi_3}},
\end{align}
where:
\begin{align*}
\Theta_1 &= \frac{i T \alpha_1 (\sigma_{22} + 2\sigma_{33}) + \beta_1 \left[%
T \sigma_{22} + 4\beta_1 (\delta_1 + i\chi_1) \right] + 4\alpha_1^2
(\delta_1 + i\chi_1)}{2(\alpha_1^2 + \beta_1^2)}, \\
\Theta_2 &= \frac{i T \alpha_1 (\sigma_{11} + 2\sigma_{33}) + \beta_1 \left[%
T \sigma_{11} + 4\beta_1 (\delta_1 + i\chi_1) \right] + 4\alpha_1^2
(\delta_1 + i\chi_1)}{2(\alpha_1^2 + \beta_1^2)}, \\
\Xi_1 &= \frac{i}{2}\left[4x(\alpha_1 + i\beta_1) + \frac{T(\sigma_{11} +
\sigma_{22})}{\alpha_1 + i\beta_1} + \frac{T\sigma_{33}}{\alpha_1 - i\beta_1}
\right], \\
\Xi_2 &= 2x(i\alpha_1 + \beta_1) + 4\delta_1 + \frac{iT\sigma_{22}}{%
2(\alpha_1 - i\beta_1)} + \frac{iT(\sigma_{11} + \sigma_{33})}{2(\alpha_1 +
i\beta_1)}, \\
\Xi_3 &= 2x(i\alpha_1 + \beta_1) + 4\delta_1 + \frac{iT\sigma_{11}}{%
2(\alpha_1 - i\beta_1)} + \frac{iT(\sigma_{22} + \sigma_{33})}{2(\alpha_1 +
i\beta_1)}.
\end{align*}
This explicit expression demonstrates the structure of the one-soliton
solution and reveals how each physical parameter shapes the resulting pulse
behavior in the medium.

The one-soliton solution describes a localized, stable pair of pulses in the
probe and coupling fields. Its shape and speed are controlled by the
spectral parameter $\lambda _{1}\equiv \alpha _{1}+i\beta _{1}$, where $%
\alpha _{1}$ produces the soliton's propagation speed, while $\beta _{1}$
determines the soliton's width, The population-matrix entries $\sigma _{ij}$
affect the phase dynamics through the interaction with the medium, and
parameters $\delta _{1}$ and $\chi _{1}$ provide tunable phase and position
offsets through the gauge transformation.

The solution's denominator in Eqs. (29) and (30) involves the combination of
three exponentials, representing the interference contributions from the
three-level structure of the atomic medium. These exponential terms ensure
that the soliton maintains the coherence and avoids spreading, consistently
with the integrable nature of the system.

\subsection{The derivation of the two-soliton solution}

The two-soliton solution of the MB system, produced by the gauge
transformation, is given by the general formula:
\begin{equation}
q_{k}^{[2]}(x,T)=-\sum_{j=1}^{2}2i(\lambda _{j}-\lambda _{j}^{\ast })\frac{%
(\Phi _{j}\Phi _{j}^{\dagger })_{3k}}{\Phi _{j}^{\dagger }\Phi _{j}},\quad
k=1,2,
\end{equation}%
where each auxiliary vector is constructed as
\begin{equation}
\Phi _{j}=\Psi ^{\lbrack 0]}(x,T;\lambda _{j})\mathbf{v}_{j},\quad \text{with%
}\quad \mathbf{v}_{j}=(\varepsilon _{1j},\varepsilon _{2j},1)^{\mathrm{tr}%
},\quad j=1,2.  \notag
\end{equation}%
Repeating the same approach and using the one-soliton solution, we construct
the explicit form of the two-soliton solution as
\begin{align}
q_{1}^{[2]}(x,T)& =-\frac{4\sqrt{2}\,\beta _{1}\,\varepsilon _{11}\,\exp
^{\Xi _{12}}}{2\sum\limits_{k=1}^{3}\exp \left( \Omega _{k}^{(1)}\right) }-%
\frac{4\sqrt{2}\,\beta _{2}\,\varepsilon _{12}\,\exp ^{\Xi _{22}}}{%
2\sum\limits_{k=1}^{3}\exp \left( \Omega _{k}^{(2)}\right) }, \\[1.5ex]
q_{2}^{[2]}(x,T)& =-\frac{4\sqrt{2}\,\beta _{1}\,\varepsilon _{21}\,\exp
^{\Xi _{12^{\prime }}}}{2\sum\limits_{k=1}^{3}\exp \left( \Omega
_{k}^{(1)}\right) }-\frac{4\sqrt{2}\,\beta _{2}\,\varepsilon _{22}\,\exp
^{\Xi _{22^{\prime }}}}{2\sum\limits_{k=1}^{3}\exp \left( \Omega
_{k}^{(2)}\right) },
\end{align}
where
\begin{align*}
\Xi _{12}& =\frac{iT\alpha _{1}(\sigma _{22}+2\sigma _{33})+\beta _{1}\left(
T\sigma _{22}+4\beta _{1}(\delta _{1}+i\chi _{1})\right) +4\alpha
_{1}^{2}(\delta _{1}+i\chi _{1})}{2(\alpha _{1}^{2}+\beta _{1}^{2})}, \\%
[1.5ex]
\Xi _{12^{\prime }}& =\frac{iT\alpha _{1}(\sigma _{11}+2\sigma _{33})+\beta
_{1}\left( T\sigma _{11}+4\beta _{1}(\delta _{1}+i\chi _{1})\right) +4\alpha
_{1}^{2}(\delta _{1}+i\chi _{1})}{2(\alpha _{1}^{2}+\beta _{1}^{2})},
\end{align*}%
and
\begin{align*}
\Xi _{22}& =\frac{1}{2}\left( 4\delta _{2}+\frac{i\left[ T(\alpha
_{2}+i\beta _{2})\sigma _{33}+(\alpha _{2}-i\beta _{2})\left( T(\sigma
_{22}+\sigma _{33})+4(\alpha _{2}+i\beta _{2})\chi _{2}\right) \right] }{%
(\alpha _{2}-i\beta _{2})(\alpha _{2}+i\beta _{2})}\right) , \\[1.5ex]
\Xi _{22^{\prime }}& =\frac{1}{2}\left( 4\delta _{2}+\frac{i\left[ T(\alpha
_{2}+i\beta _{2})\sigma _{33}+(\alpha _{2}-i\beta _{2})\left( T(\sigma
_{11}+\sigma _{33})+4(\alpha _{2}+i\beta _{2})\chi _{2}\right) \right] }{%
(\alpha _{2}-i\beta _{2})(\alpha _{2}+i\beta _{2})}\right) ,
\end{align*}%
with the arguments of the denominator exponentials (for $j=1,2$)

\begin{align*}
\Omega _{1}^{(j)}& =\frac{i}{2}\left( 4x(\alpha _{j}+i\beta _{j})+\frac{%
T(\sigma _{11}+\sigma _{22})}{\alpha _{j}+i\beta _{j}}+\frac{T\sigma _{33}}{%
\alpha _{j}-i\beta _{j}}\right) , \\[1.5ex]
\Omega _{2}^{(j)}& =2x(i\alpha _{j}+\beta _{j})+4\delta _{j}+\frac{iT\sigma
_{22}}{2(\alpha _{j}-i\beta _{j})}+\frac{iT(\sigma _{11}+\sigma _{33})}{%
2(\alpha _{j}+i\beta _{j})}, \\[1.5ex]
\Omega _{3}^{(j)}& =2x(i\alpha _{j}+\beta _{j})+4\delta _{j}+\frac{iT\sigma
_{11}}{2(\alpha _{j}-i\beta _{j})}+\frac{iT(\sigma _{22}+\sigma _{33})}{%
2(\alpha _{j}+i\beta _{j})}.
\end{align*}

The two-soliton solution reflects the interplay of two localized wave
packets co-evolving in the $\Lambda $-type EIT medium. Each soliton is
produced by its spectral parameter $\lambda _{j}=\alpha _{j}+i\beta _{j}$,
which determines its speed and shape. Expressions (32) and (33) clearly
distinguish the contributions of each soliton to both fields $q_{1}$ and $%
q_{2}$, while the the exponentials in the shared denominators account for
their interference. The denominator includes population coefficients $\sigma
_{ij}$, which mediate the soliton-soliton interaction through the atomic
coherence.

This solution represents the situation in which two light pulses propagate
with different group velocities, embedded in the probe and coupling beams.
The variation of the amplitude and phase results from their mutual impact,
which is enhanced by the atomic ensemble. The interference terms in the
denominator exhibit the beat-like structures resulting in intensity
oscillations and the transient trapping. Such effects are hallmarks of the
coherent population trapping and nonlinear pulse steering, which are
inherent in the multi-soliton EIT dynamics.

\subsection{The explicit form of the three-soliton solution}

Similarly, by applying the gauge-transformation iteration to the two-soliton
solution, we obtain the three-soliton solution of the MB equation in a
sufficiently compact form as
\begin{align}
q_{1}^{[3]}(x,T)=& -\frac{4\sqrt{2}\,\beta _{1}\,\varepsilon _{11}\cdot \exp
\left( \frac{iT\alpha _{1}(\sigma _{22}+2\sigma _{33})+\beta _{1}(T\sigma
_{22}+4\beta _{1}(\delta _{1}+i\chi _{1}))+4\alpha _{1}^{2}(\delta
_{1}+i\chi _{1})}{2(\alpha _{1}^{2}+\beta _{1}^{2})}\right) }{%
2(D_{1}^{(1)}+D_{2}^{(1)}+D_{3}^{(1)})}  \notag \\
& -\sum_{j=2}^{3}\frac{4\sqrt{2}\,\beta _{j}\,\varepsilon _{1j}\cdot \exp
\left( \frac{1}{2}\left[ 4\delta _{j}+\frac{i\left[ T(\alpha _{j}+i\beta
_{j})\sigma _{33}+(\alpha _{j}-i\beta _{j})(T\sigma _{22}+t\sigma
_{33}+4(\alpha _{j}+i\beta _{j})\chi _{j})\right] }{(\alpha _{j}-i\beta
_{j})(\alpha _{j}+i\beta _{j})}\right] \right) }{%
2(D_{1}^{(j)}+D_{2}^{(j)}+D_{3}^{(j)})},
\end{align}%
and

\begin{align}
q_{2}^{[3]}(x,t)=& -\frac{4\sqrt{2}\,\beta _{1}\,\varepsilon _{21}\cdot \exp
\left( \frac{iT\alpha _{1}(\sigma _{11}+2\sigma _{33})+\beta _{1}(T\sigma
_{11}+4\beta _{1}(\delta _{1}+i\chi _{1}))+4\alpha _{1}^{2}(\delta
_{1}+i\chi _{1})}{2(\alpha _{1}^{2}+\beta _{1}^{2})}\right) }{%
2(D_{1}^{(1)}+D_{2}^{(1)}+D_{3}^{(1)})}  \notag \\
& -\sum_{j=2}^{3}\frac{4\sqrt{2}\,\beta _{j}\,\varepsilon _{2i}\cdot \exp
\left( \frac{1}{2}\left[ 4\delta _{j}+\frac{i\left[ T(\alpha _{j}+i\beta
_{j})\sigma _{33}+(\alpha _{j}-i\beta _{j})(T\sigma _{11}+t\sigma
_{33}+4(\alpha _{j}+i\beta _{j})\chi _{j})\right] }{(\alpha _{j}-i\beta
_{j})(\alpha _{j}+i\beta _{j})}\right] \right) }{%
2(D_{1}^{(j)}+D_{2}^{(j)}+D_{3}^{(j)})},
\end{align}%
with the denominator terms for $j=1,2,3$ being
\begin{align*}
D_{1}^{(j)}& =\exp \left( \frac{i}{2}\left[ 4x(\alpha _{j}+i\beta _{j})+%
\frac{T(\sigma _{11}+\sigma _{22})}{\alpha _{j}+i\beta _{j}}+\frac{T\sigma
_{33}}{\alpha _{j}-i\beta _{j}}\right] \right) , \\
D_{2}^{(j)}& =\exp \left( 2x(i\alpha _{j}+\beta _{j})+4\delta _{j}+\frac{%
iT\sigma _{22}}{2(\alpha _{j}-i\beta _{j})}+\frac{iT(\sigma _{11}+\sigma
_{33})}{2(\alpha _{j}+i\beta _{j})}\right) , \\
D_{3}^{(j)}& =\exp \left( 2x(i\alpha _{j}+\beta _{j})+4\delta _{j}+\frac{%
iT\sigma _{11}}{2(\alpha _{j}-i\beta _{j})}+\frac{iT(\sigma _{22}+\sigma
_{33})}{2(\alpha _{j}+i\beta _{j})}\right) .
\end{align*}%
The three-soliton solution introduces a diverse field structures arising
from the interaction of the three coherent pulses. The polarization vectors $%
\mathbf{v}_{j}=(\varepsilon _{1j},\varepsilon _{2j},1)^{T}$ and gauge
parameters $\delta _{j},\chi _{j}$ determine the individual soliton
contributions to each term in the solution, providing the precise control
over the relative phase and position. The exponential phase terms in the
denominators represent the spatiotemporal shifts of the solitons caused by
the cross-modulation.

These multi-soliton arrangements pave the way for soliton-based
multiplexing, which is essential to quantum switching and photonic logic
because it allows multiple data channels to propagate concurrently and
interact coherently without degrading.

\subsection{The construction of the four-soliton solution}

Next, we construct the four-soliton solution of the integrable MB equations
by means of the gauge iterative method:
\begin{align}
q_{1}^{[4]}(x,T)=& -\frac{4\sqrt{2}\,\beta _{1}\,\varepsilon _{11}\cdot \exp
\left( \frac{iT\alpha _{1}(\sigma _{22}+2\sigma _{33})+\beta _{1}\left(
T\sigma _{22}+4\beta _{1}(\delta _{1}+i\chi _{1})\right) +4\alpha
_{1}^{2}(\delta _{1}+i\chi _{1})}{2(\alpha _{1}^{2}+\beta _{1}^{2})}\right)
}{2\left( D_{1}^{(1)}+D_{2}^{(1)}+D_{3}^{(1)}\right) }  \notag \\
& -\sum_{j=2}^{4}\frac{4\sqrt{2}\,\beta _{j}\,\varepsilon _{1j}\cdot \exp
\left( \frac{1}{2}\left[ 4\delta _{j}+\frac{i\left[ T(\alpha _{j}+i\beta
_{j})\sigma _{33}+(\alpha _{j}-i\beta _{j})\left( T\sigma _{22}+t\sigma
_{33}+4(\alpha _{j}+i\beta _{j})\chi _{j}\right) \right] }{(\alpha
_{j}-i\beta _{j})(\alpha _{j}+i\beta _{j})}\right] \right) }{2\left(
D_{1}^{(j)}+D_{2}^{(j)}+D_{3}^{(j)}\right) },
\end{align}%
and
\begin{align}
q_{2}^{[4]}(x,T)=& -\frac{4\sqrt{2}\,\beta _{1}\,\varepsilon _{21}\cdot \exp
\left( \frac{iT\alpha _{1}(\sigma _{11}+2\sigma _{33})+\beta _{1}\left(
T\sigma _{11}+4\beta _{1}(\delta _{1}+i\chi _{1})\right) +4\alpha
_{1}^{2}(\delta _{1}+i\chi _{1})}{2(\alpha _{1}^{2}+\beta _{1}^{2})}\right)
}{2\left( \mathcal{D}_{1}^{(1)}+\mathcal{D}_{2}^{(1)}+\mathcal{D}%
_{3}^{(1)}\right) }  \notag \\
& -\sum_{j=2}^{4}\frac{4\sqrt{2}\,\beta _{j}\,\varepsilon _{2j}\cdot \exp
\left( \frac{1}{2}\left[ 4\delta _{j}+\frac{i\left[ T(\alpha _{j}+i\beta
_{j})\sigma _{33}+(\alpha _{j}-i\beta _{j})\left( T\sigma _{11}+t\sigma
_{33}+4(\alpha _{j}+i\beta _{j})\chi _{j}\right) \right] }{(\alpha
_{j}-i\beta _{j})(\alpha _{j}+i\beta _{j})}\right] \right) }{2\left(
\mathcal{D}_{1}^{(j)}+\mathcal{D}_{2}^{(j)}+\mathcal{D}_{3}^{(j)}\right) },
\end{align}%
with the following common denominator terms for $j=1,2,3,4$:
\begin{align*}
\mathcal{D}_{1}^{(j)}& =\exp \left( \frac{i}{2}\left[ 4x(\alpha _{j}+i\beta
_{j})+\frac{T(\sigma _{11}+\sigma _{22})}{\alpha _{j}+i\beta _{j}}+\frac{%
T\sigma _{33}}{\alpha _{j}-i\beta _{j}}\right] \right) , \\
\mathcal{D}_{2}^{(j)}& =\exp \left( 2x(i\alpha _{j}+\beta _{j})+4\delta _{j}+%
\frac{iT\sigma _{22}}{2(\alpha _{j}-i\beta _{j})}+\frac{iT(\sigma
_{11}+\sigma _{33})}{2(\alpha _{j}+i\beta _{j})}\right) , \\
\mathcal{D}_{3}^{(j)}& =\exp \left( 2x(i\alpha _{j}+\beta _{j})+4\delta _{j}+%
\frac{iT\sigma _{11}}{2(\alpha _{j}-i\beta _{j})}+\frac{iT(\sigma
_{22}+\sigma _{33})}{2(\alpha _{j}+i\beta _{j})}\right) .
\end{align*}

The four-soliton solution represents the most intricate one considered here,
encapsulating the interactions of four independently tunable soliton
components. The algebraic structure of the solution underscores the additive
nature of the coherent excitations in the integrable MB system, while the
denominators again enforce the nonlinear coupling mediated by the system's
three-level structure. The population terms $\sigma _{11},\sigma
_{22},\sigma _{33}$ now have a substantial impact on how the solitons split
the energy between fields and sustain coherence, highlighting the prominent
significance of the atomic coherence.

\section{Discussion of the soliton states and their interactions}

\subsection{The interpretation of the one-soliton profile}

Figure 1 displays the one-soliton solution, given by Eqs.(29) and (30), for
the integrable MB system of Eqs. (5)-(7). The distinctive temporal asymmetry
between the components $q_{1}(x,t)$ and $q_{2}(x,t)$ in the one-soliton
solution is directly caused by the initial density matrix configuration and
the complicated phase structure of the solution. We consider the initial
condition where the population is fully placed in the ground state $%
|1\rangle $ ($\sigma _{11}=1$, $\sigma _{22}=\sigma _{33}=0$), and the
coupling amplitudes are $\varepsilon _{11}=0.5$, $\varepsilon _{21}=\sqrt{%
1-\varepsilon _{11}^{2}}\approx \allowbreak 0.866$. In this configuration, the probing pulse $q_{1}$ emerges and propagates in
the negative time direction, while the pulse $q_{2}$ evolves in the positive
time direction. This temporal asymmetry is caused by the different phase
contributions from the $\sigma _{ij}$ factors in each component, which leads
to the group delay between the pulses. 

The spectral parameter $\lambda_{1} = \alpha_{1} + i\beta_{1}$ controls the soliton kinematics, where the real part $\alpha_{1} = 0.4$ sets the group velocity and the imaginary part $\beta_{1} = 0.7$ determines the temporal confinement of the pulse. The small gauge offsets $\delta_{1} = 0.02$ and $\chi_{1} = 0.03$ introduce slight temporal and phase displacements between the probe and coupling components in Fig.~1. Hence, while the polarization and population parameters define the energy distribution, the spectral and gauge parameters tune the width, velocity, and phase evolution of the soliton during propagation.

\begin{figure}[H]
\centering
\includegraphics[width=10cm]{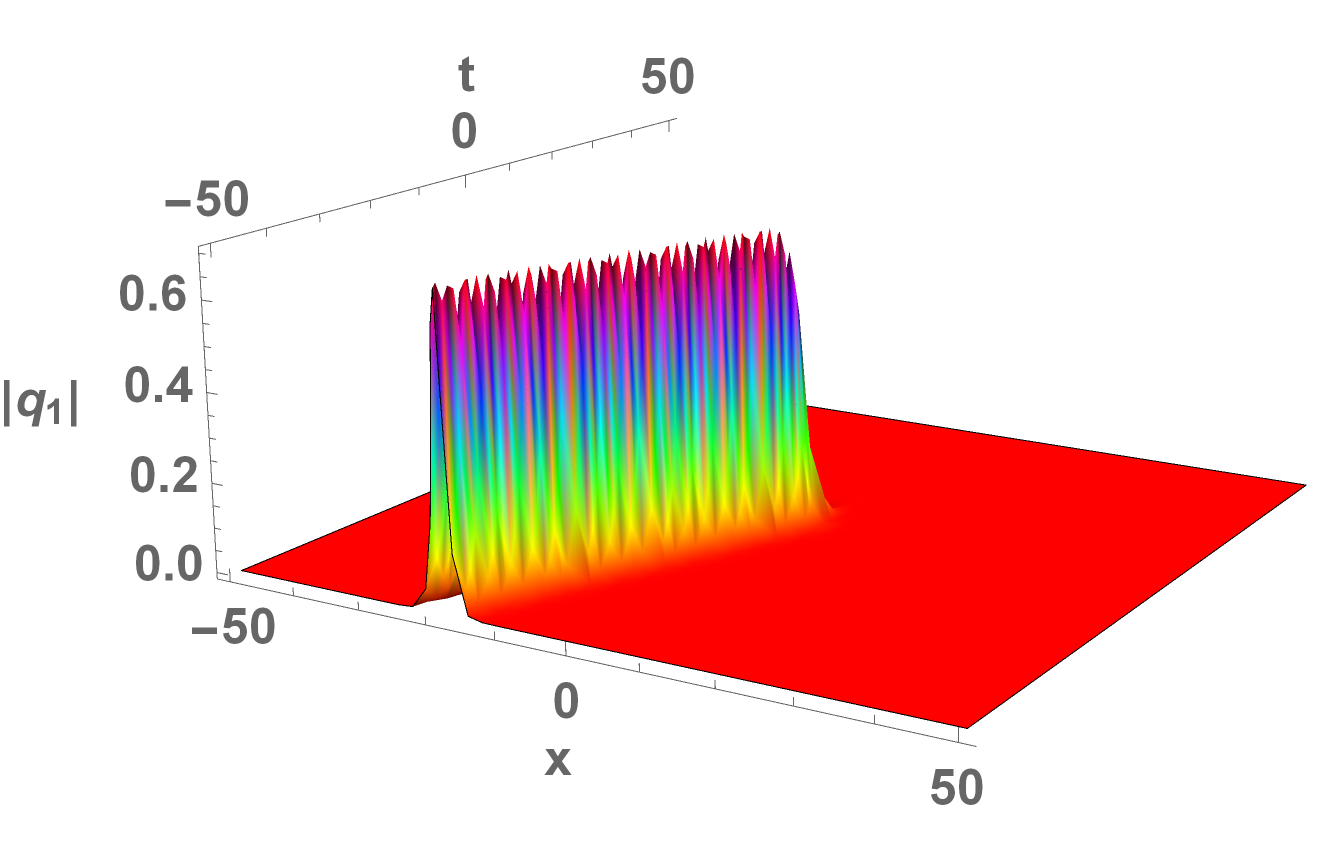}
\includegraphics[width=10cm]{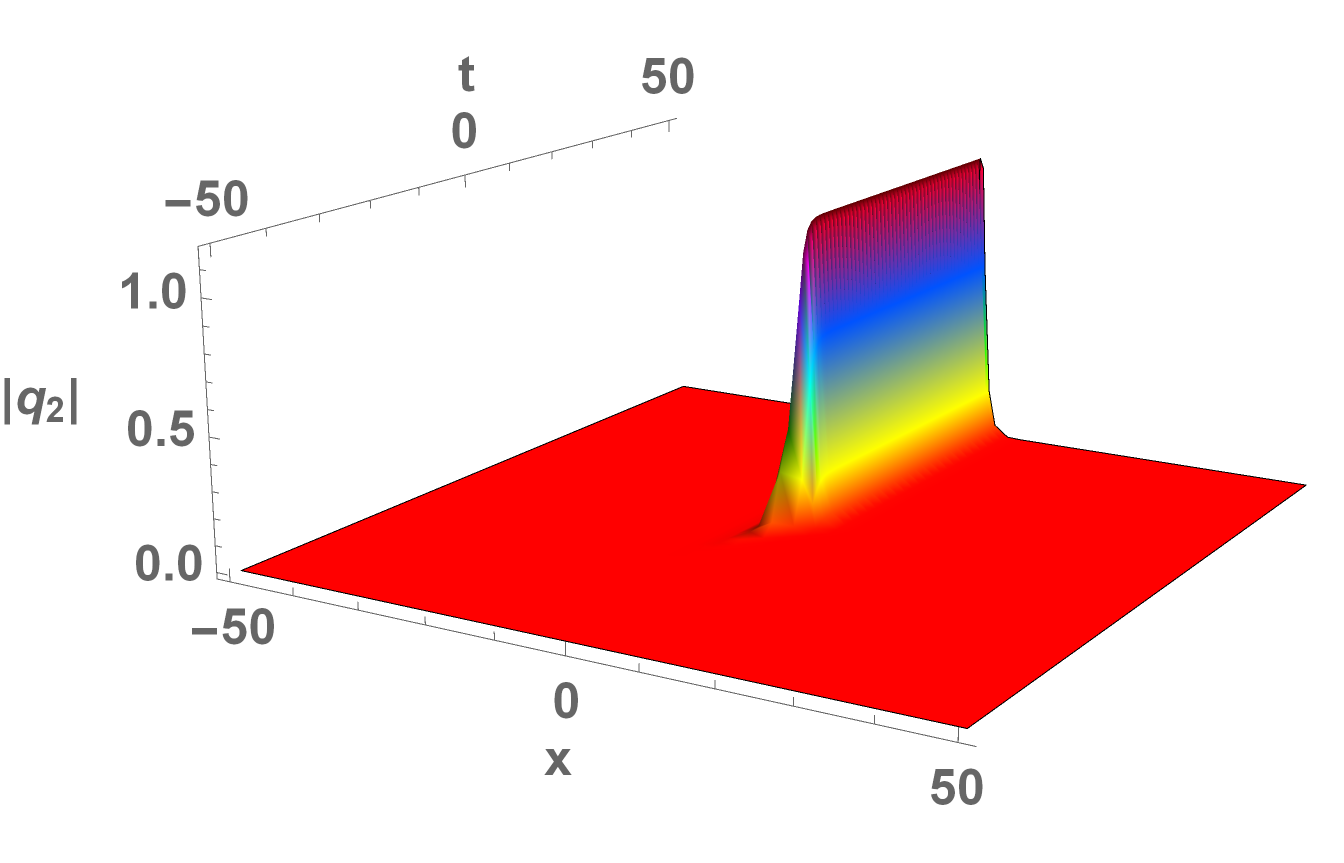}
\caption{The one-soliton solution given by Eqs. (29) and (30) for the
parameter set $\protect\alpha _{1}=0.4$, $\protect\beta _{1}=0.7$, $\protect%
\delta _{1}=0.02$, $\protect\chi _{1}=0.03$, $\protect\varepsilon _{11}=0.5$%
, $\protect\varepsilon _{21}=\protect\sqrt{1-\protect\varepsilon _{11}^{2}}%
\approx 0.866$, with the initial density-matrix centies $\protect\sigma %
_{11}=1$, $\protect\sigma _{22}=\protect\sigma _{33}=0$.}
\label{fig:onesoliton}
\end{figure}

The soliton associated with $q_{1}$ can be physically understood as entering
the medium from the negative time domain and getting trapped by the coherent
interaction with atoms. This stored excitation is then re-emitted as the
soliton in $q_{2}$, that is generated and travels forward in time. The pulse
trapping and re-emission is observed in Fig.~\ref{fig:onesoliton}, being a
soliton counterpart of the behavior of stationary light pulses demonstrated
in Refs. \cite{Bajcsy2003,Lukin2003,Chaneliere2005}.

\subsection{The interpretation of the two-soliton pulse propagation}

\begin{figure}[H]
\centering
\includegraphics[width=10cm]{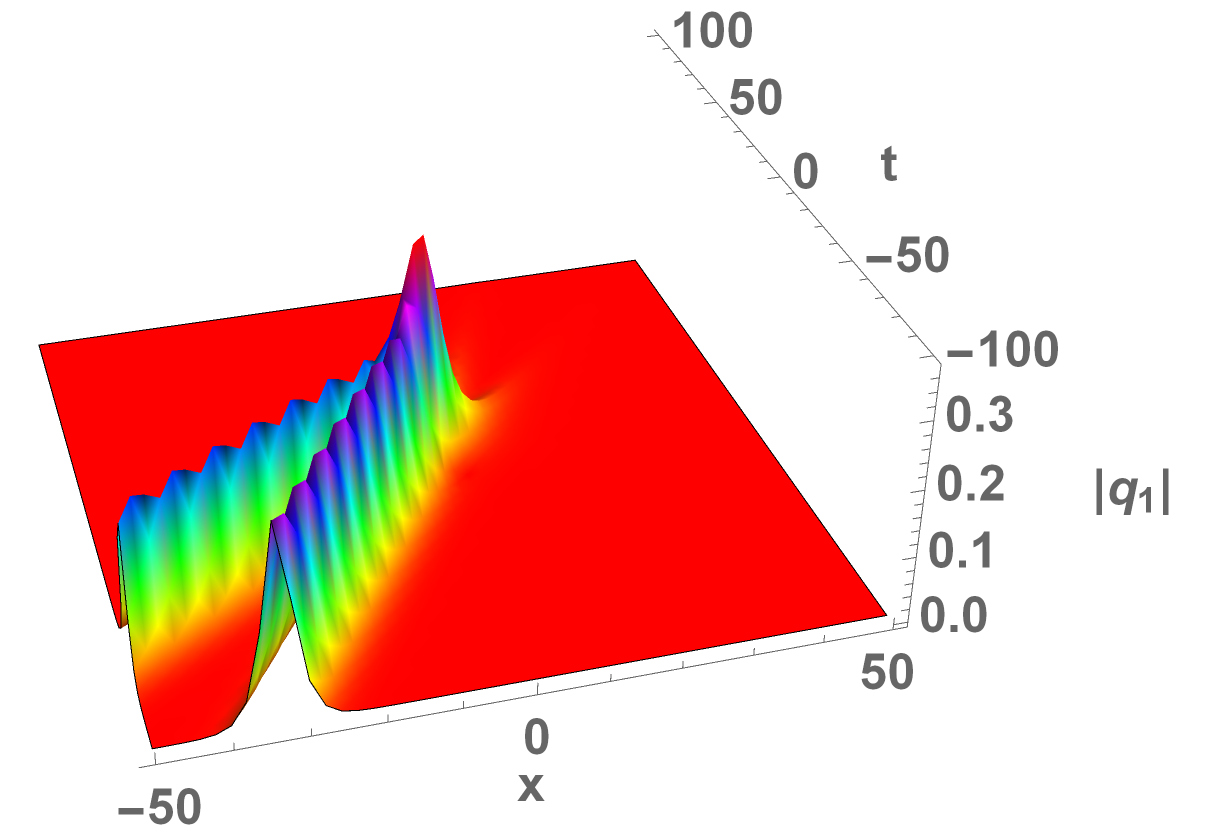}
\includegraphics[width=10cm]{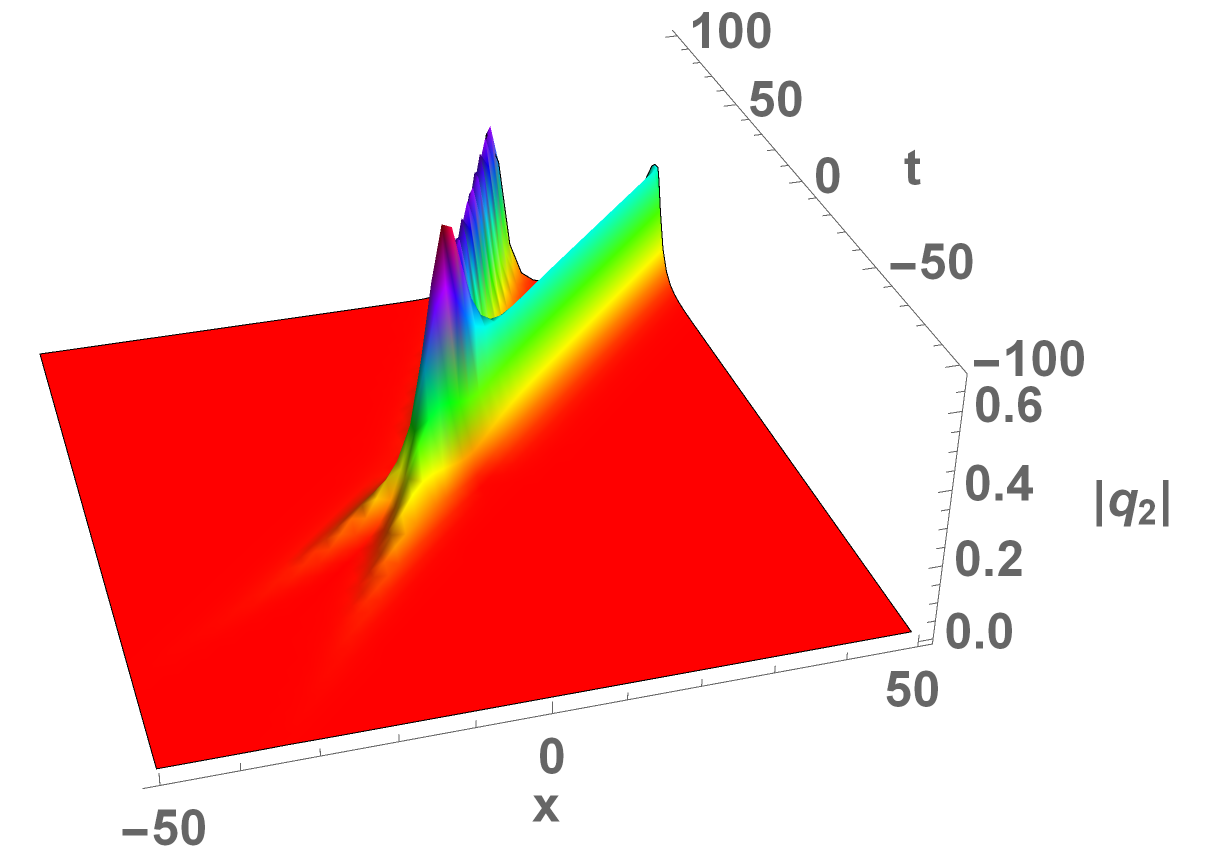}
\caption{The two-soliton solution given by Eqs.(32) and (33) for the
parameter set: $\protect\alpha _{1}=0.4$, $\protect\alpha _{2}=0.6$, $%
\protect\beta _{1}=0.2$, $\protect\beta _{2}=0.3$, $\protect\delta _{1}=0.3$%
, $\protect\delta _{2}=0.2$, $\protect\chi _{1}=0.03$, $\protect\chi %
_{2}=0.01$, $\protect\varepsilon _{11}=0.6$, $\protect\varepsilon _{12}=0.5$%
, with the initial density-matrix values $\protect\sigma _{11}=0.6$, $%
\protect\sigma _{22}=0.4$, $\protect\sigma _{33}=0$.}
\label{fig:twosoliton}
\end{figure}

Figure~\ref{fig:twosoliton} displays the coherent population dynamics
primarily distributed between the two ground states for the initial
density-matrix values $\sigma _{11}=0.6$, $\sigma _{22}=0.4$, and $\sigma
_{33}=0$, for the two-soliton solution given by Eqs. (32) and (33). The
absence of the population in the excited state ensures low absorption and
provides a suitable medium for the production of the soliton through the
destructive quantum interference. The spectral parameters $\alpha_{1,2}$ and $\beta_{1,2}$ dictate the relative velocities and temporal widths of the two pulses, while the gauge parameters $\delta_{i}$ and $\chi_{i}$ ($i=1,2$) introduce phase shifts that generate the oscillatory interference observed in Fig.~2. The chosen values $\alpha_{1}=0.4$, $\alpha_{2}=0.6$, $\beta_{1}=0.2$, $\beta_{2}=0.3$, together with $\delta_{1}=0.3$, $\delta_{2}=0.2$, $\chi_{1}=0.03$, and $\chi_{2}=0.01$, demonstrate how differential group velocities and small phase offsets produce beating-like intensity oscillations and transient energy trapping between the soliton components.  The successful trapping of the
probe pulses and their subsequent release in the form of coupling-field
pulses demonstrates a partial storage and retrieval mechanism integrated
into EIT, cf. Refs. \cite{Lukin2003, Hedges2010, Simon2010}.

\subsection{The elastic collisional behavior of the three-soliton}

\begin{figure}[H]
\centering
\includegraphics[width=10cm]{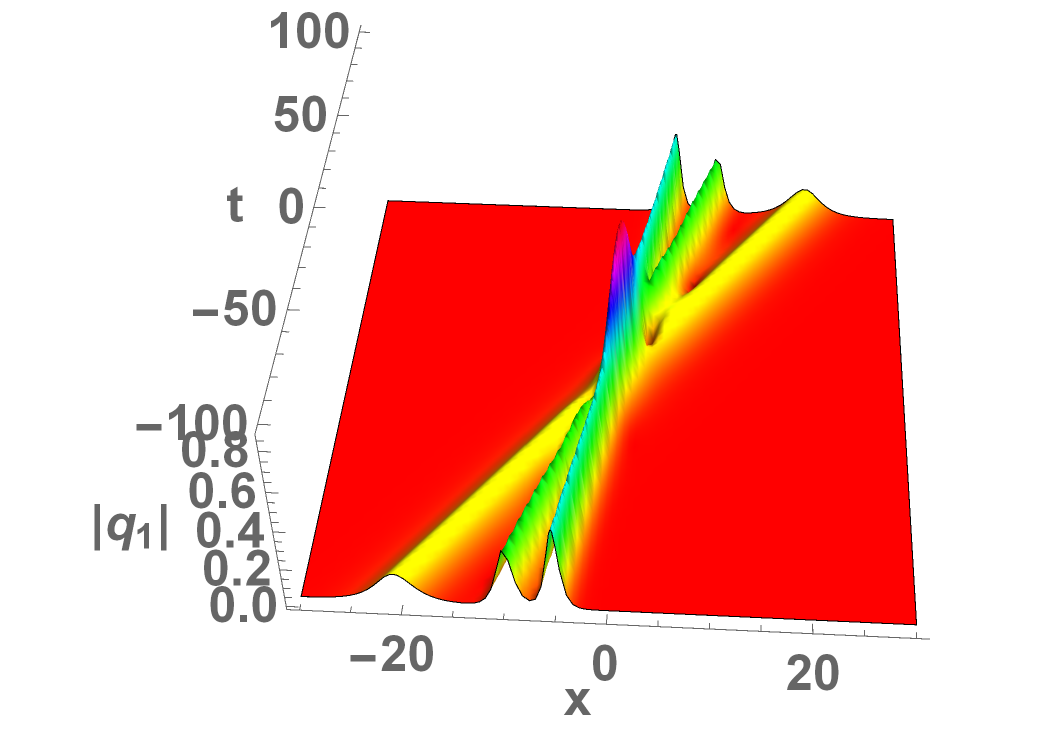}
\includegraphics[width=10cm]{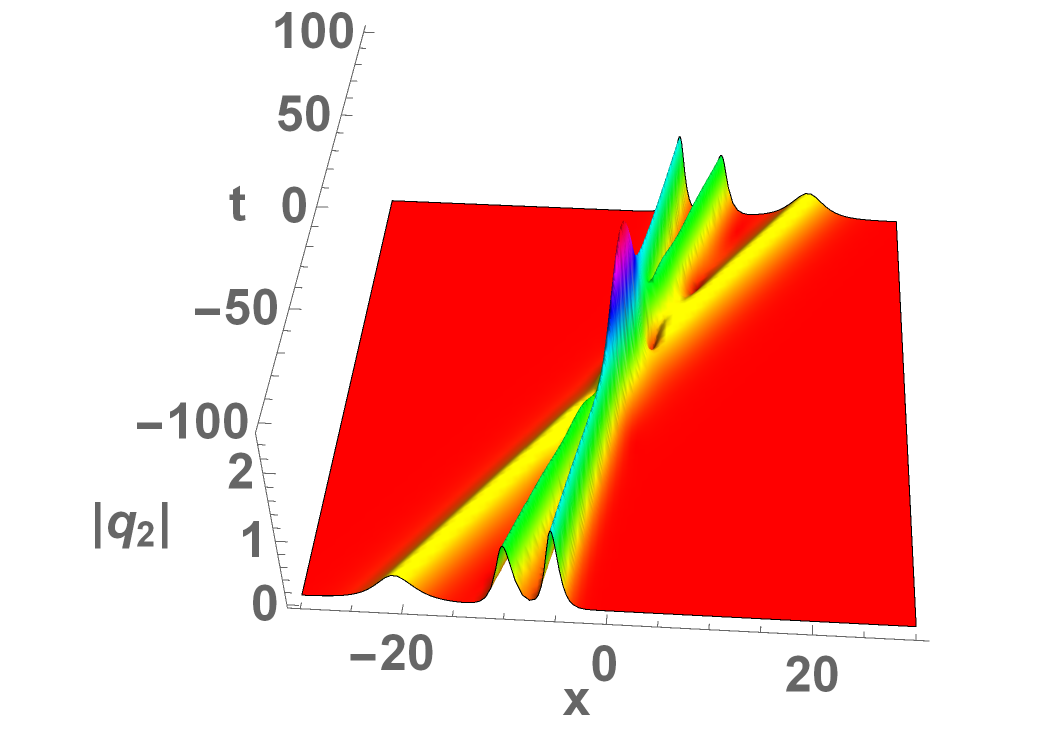}
\caption{The three-soliton solution given by Eqs. (34) and (35) for the
parameter set $\protect\alpha _{1}=0.4$, $\protect\alpha _{2}=0.2$, $\protect%
\alpha _{3}=0.5$, $\protect\beta _{1}=0.3$, $\protect\beta _{2}=0.7$, $%
\protect\beta _{3}=0.9$, $\protect\delta _{1}=0.2$, $\protect\delta _{2}=0.3$%
, $\protect\delta _{3}=0.5$, $\protect\chi _{1}=0.03$, $\protect\chi %
_{2}=0.05$, $\protect\chi _{3}=0.07$, with the coupling coefficients $%
\protect\varepsilon _{11}=0.35$, $\protect\varepsilon _{12}=0.31$, $\protect%
\varepsilon _{13}=0.331$ and the initial values of the density matrix $%
\protect\sigma _{11}=0.4$, $\protect\sigma _{22}=0.4$, $\protect\sigma %
_{33}=0.2$.}
\label{fig:three-soliton}
\end{figure}

The pulse dynamics of the three-soliton solution of the MB system is
featured by localized structures in $|q_{1}^{[3]}(x,T)|$ and $%
|q_{2}^{[3]}(x,T)|$, as shown in Fig. \ref{fig:three-soliton}. The initial
values of the density matrix chosen here, $\sigma _{11}=0.4$, $\sigma
_{22}=0.4$, and $\sigma _{33}=0.2$, demonstrate that the atomic population
is nearly evenly distributed in the lower states, with a very small share in
the excited state. In this configuration, EIT allows several solitons to
move across the atomic medium. The spectral parameters $\alpha_{j}$ and $\beta_{j}$ ($j=1,2,3$) determine the distinct velocities and compressions of each soliton. Larger $\beta_{3}=0.9$ yields a narrower and faster pulse, whereas smaller $\beta_{1}=0.3$ produces a broader, slower one. The gauge parameters $\delta_{j}$ and $\chi_{j}$ ($\delta_{1,2,3}=0.2,0.3,0.5$; $\chi_{1,2,3}=0.03,0.05,0.07$) set the relative temporal offsets and fine-tune the phase coherence among the three interacting pulses. The coupling coefficients $\varepsilon_{11}=0.35$, $\varepsilon_{12}=0.31$, and $\varepsilon_{13}=0.331$ modulate the amplitude ratio between the probe and coupling fields.  These parameter combinations reproduce the elastic collision behaviour and coherent population redistribution shown in Fig.~3.

In contrast to the one- and two-soliton solutions, the three-soliton
configurations displayed in Fig.~\ref{fig:three-soliton} exhibit the
continuous and forward-directed evolution of all pulses without any obvious
indication of field trapping. The solitons in $|q_{1}^{[3]}|$ and $%
|q_{2}^{[3]}|$ freely travel throughout the medium and produce complex wave
interactions due to the improved nonlinear coupling. This fact highlights
the connection between the optical-field coherence and atomic-population
distribution in the multi-soliton regimes, demonstrating that the system
transitions from the trapping-dominant scenario to the soliton transmission, cf. Ref. \cite{Duan2001}.

\subsection{The four-soliton pulse propagation}

\begin{figure}[H]
\centering
\includegraphics[width=8cm]{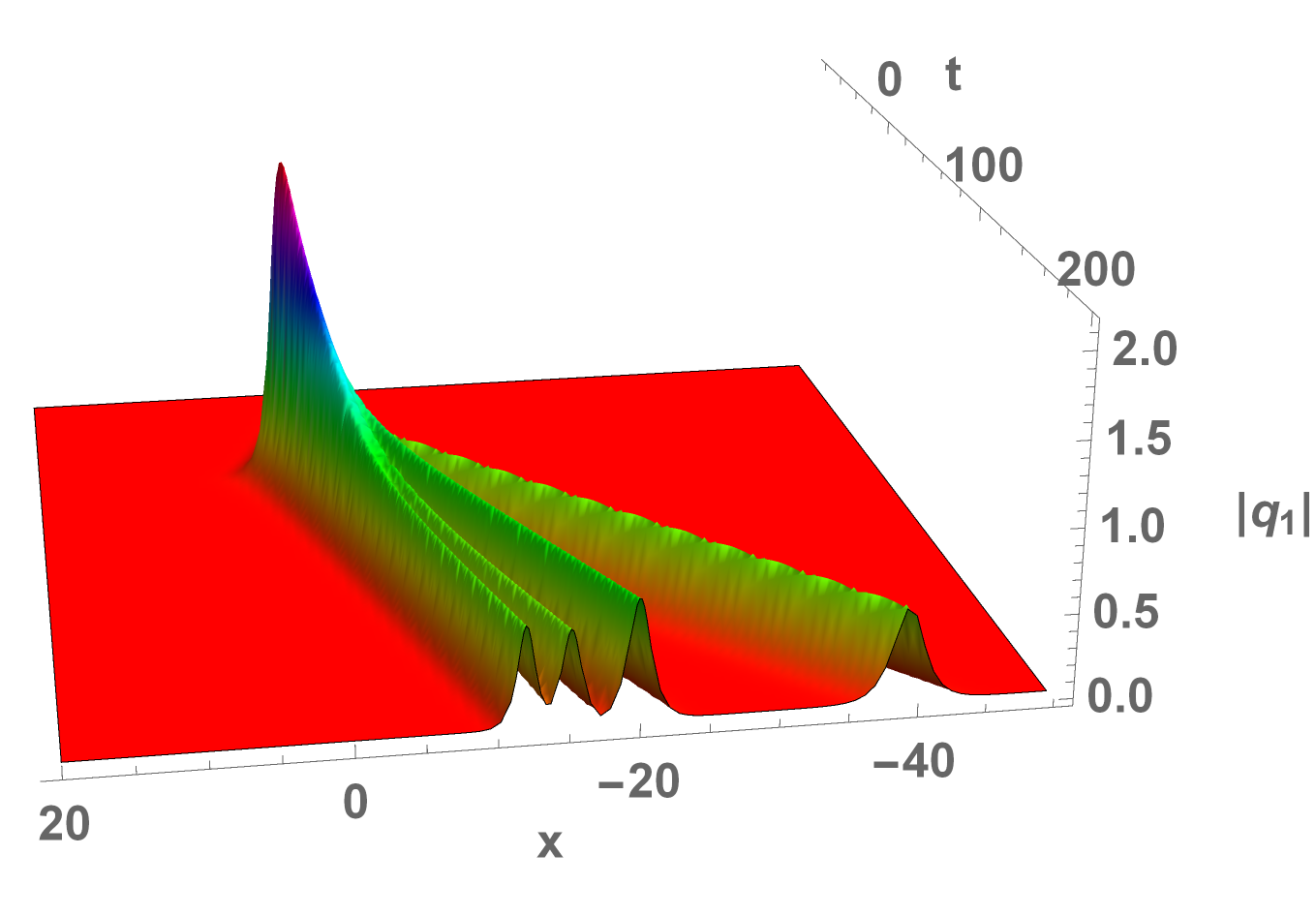}
\includegraphics[width=8cm]{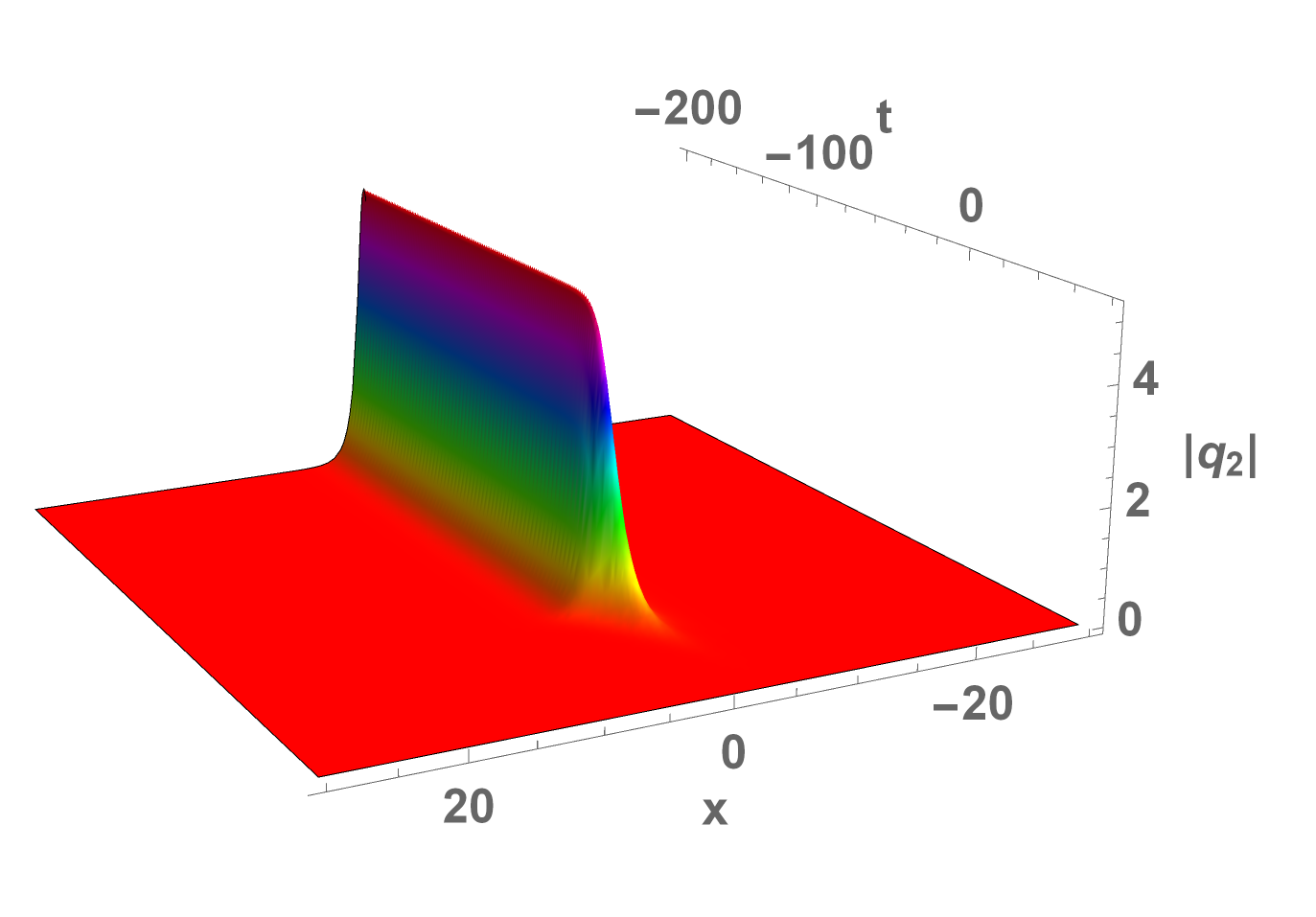}
\caption{The four-soliton solution of Eqs. (36) and (37) for the parameter
set $\protect\alpha _{1}=0.04$, $\protect\alpha _{2}=0.01$, $\protect\alpha %
_{3}=0.02$, $\protect\alpha _{4}=0.03$; $\protect\beta _{1}=0.8$, $\protect%
\beta _{2}=0.7$, $\protect\beta _{3}=0.9$, $\protect\beta _{4}=0.5$; $%
\protect\delta _{1}=0.02$, $\protect\delta _{2}=0.03$, $\protect\delta %
_{3}=0.05$, $\protect\delta _{4}=0.06$; $\protect\chi _{1}=0.03$, $\protect%
\chi _{2}=0.05$, $\protect\chi _{3}=0.07$, $\protect\chi _{4}=0.06$, with
the initial density-matrix values $\protect\sigma _{11}=0.2$, $\protect%
\sigma _{22}=0.4$, $\protect\sigma _{33}=0.4$; $\protect\varepsilon %
_{11}=0.35$, $\protect\varepsilon _{12}=0.51$, $\protect\varepsilon %
_{13}=0.331$.}
\label{fig:four-soliton}
\end{figure}

The four-soliton solution given by Eqs. (36) and (37) is displayed in
Fig.~\ref{fig:four-soliton}. In this case, fields $q_{1}$ and $q_{2}$
exhibit a pronounced asymmetry. The probe field $q_{1}$ displays four
well-separated soliton pulses, while the coupling field $q_{2}$ carries just
one dominant soliton. In particular, the values $\sigma _{11}=0.2$, $\sigma
_{22}=0.4$, and $\sigma _{33}=0.4$ restrict the coupling dynamics to a more
confined structure, while allowing the medium to support numerous nonlinear
excitations in the probe channel. The sequence of small $\alpha_{j}$ values ($0.01$–$0.04$) and comparatively large $\beta_{j}$ values ($0.5$–$0.9$) produces strong temporal compression and well-separated localized peaks in the probe field. The gradual increase of the gauge parameters $\delta_{j}$ and $\chi_{j}$ ($0.02 \rightarrow 0.06$ and $0.03 \rightarrow 0.07$) introduces ordered phase delays that generate the pulse train observed in Fig.~4. These spectral and gauge variations, together with fixed polarization and population parameters, confirm that the multi-peak structure and asymmetric energy distribution can be precisely engineered by tuning $\alpha_{j}$, $\beta_{j}$, $\delta_{j}$, and $\chi_{j}$. This configuration increases the possibility of storing
several information channels in a single atomic ensemble, by enabling the
selective and coherent transmission of energy from the coupling field to
numerous probing solitons \cite{Pezze2018}.

The presence of many compressed solitons in the probing field is a certain
manifestation of the soliton-induced slow light, where the effective group
velocity is significantly reduced due to the temporal compression and
enhanced nonlinear interaction. This is essential for the realization of the
quantum data storage and delay lines \cite{Wu2010, Zhao2009, Ma2017}.
Furthermore, the ability to generate and send a large number of coherent
controllable soliton pulses offers possibilities for multi-qubit encoding,
quantum-memory architectures, and optical logic gates, all being crucially
important elements of quantum computing systems. Thus, the observed
asymmetry not only illustrates the system's intrinsic nonlinear dynamics,
but also its potential for the use in the frameworks of quantum
communications and computations, where tunable soliton structures can serve
as reliable non-dispersive carriers of quantum information.

As mentioned above, the imaginary part of spectral parameter $\lambda
_{j}=\alpha _{j}+i\beta _{j}$ governs the temporal breadth and localization
of the soliton, whereas the real part $\alpha _{j}$ determines its group
velocity. Narrower and more peaked solitons are associated with larger values of
$\beta _{j}$. The internal structure of the soliton is shaped by the
polarization vector $\mathbf{v}_{j}=(\varepsilon _{1j},\varepsilon _{2j},1)^{%
\mathrm{tr}}$, which also controls the amplitude and phase link between multisoliton components.

\section{Conserved quantities in multi-soliton dynamics}

A fundamental characteristic of integrable systems is the existence of an
infinite number of dynamical invariants. In this section, we analyze two key
conserved quantities for the derived N-soliton solutions: the norm ($N$),
which represents the total number of excitations (or total field intensity),
and the Hamiltonian ($H$), which represents the total energy of the system. 

\subsection{Analytical expressions for the conserved quantities}

For the integrable MB system under the consideration, the conserved
quantities can be derived from the Lax-pair structure \cite{Fleischhauer2005}. The total norm and Hamiltonian are given by
\begin{eqnarray}
N &=&\int_{-\infty }^{+\infty }\left(
|q_{1}(x,T)|^{2}+|q_{2}(x,T)|^{2}\right) dx,  \notag \\
H &=&\int_{-\infty }^{+\infty }\mathcal{E}(x,T)dx=\int_{-\infty }^{+\infty
}\left( |q_{1}|^{2}+|q_{2}|^{2}+\hbar (q_{1}^{\ast }\rho _{31}+q_{1}\rho
_{31}^{\ast }+q_{2}^{\ast }\rho _{32}+q_{2}\rho _{32}^{\ast })\right) dx,
\notag
\end{eqnarray}%
where the energy density $\mathcal{E}(x,T)$ comprises the energy stored in
the electromagnetic fields ($|q_{1}|^{2}+|q_{2}|^{2}$) and the interaction
energy between the fields and the atomic dipoles.

For the N-soliton solutions constructed by means of the gauge-transformation
method, these integrals can be evaluated analytically. The results confirm
that the total norm and Hamiltonian amount to the sum of the contributions
from individual solitons (as it should be for the integrable system):

\begin{equation}
N_{\text{total}}=\sum_{j=1}^{N}N_{j},\quad H_{\text{total}%
}=\sum_{j=1}^{N}H_{j},  \notag
\end{equation}%
where the contributions for a single soliton, characterized by the spectral
parameter $\lambda _{j}=\alpha _{j}+i\beta _{j}$ and polarization vector $%
\mathbf{v}_{j}=(\varepsilon _{1j},\varepsilon _{2j},1)^{T}$, are:

\begin{equation}
N_j = 8\beta_j (\varepsilon_{1j}^2 + \varepsilon_{2j}^2), \quad H_j =
(\alpha_j^2 - \beta_j^2) N_j.  \notag
\end{equation}

\subsection{Numerical values and the physical Interpretation}

The calculated values of these conserved quantities for the one-, two-,
three-, and four-soliton solutions at various times are presented in Table 1.

\begin{table}[htbp]
\centering
\renewcommand{\arraystretch}{1.3} 
\setlength{\tabcolsep}{10pt} 
\caption{\textbf{Conserved quantities for the multi-soliton solutions.} 
The norm ($N$) and Hamiltonian ($H$) remain constant for each soliton type at all times, 
confirming the integrability of the system.}
\vspace{0.5em}
\begin{tabular}{l c c}
\hline
\textbf{Soliton Type} & \textbf{Norm ($N$)} & \textbf{Hamiltonian ($H$)} \\
\hline
One-soliton   & 5.600  & $-1.848$  \\
Two-soliton   & 8.000  & $-0.880$  \\
Three-soliton & 13.600 & $-8.264$  \\
Four-soliton  & 20.000 & $-12.400$ \\
\hline
\end{tabular}
\label{tab:conserved}
\end{table}

The conservation of the norm $N$ and Hamiltonian $H$ in all soliton
configurations, validated numerically, ensures that the gauge-transformation
method produces consistent solutions. Importantly, the norm is additive, so
that the total excitation number for the multi-soliton states is the sum of
their constituents (e.g., $N=8.0$ for the two-soliton case, obtained from $%
N_{1}=5.6$ and $N_{2}=2.4$). This particle-like property is a characteristic
of integrable soliton dynamics and confirms that each soliton acts as an
independent carrier of the conserved excitations. The negative values of the
Hamiltonian further demonstrate that all soliton states correspond to bound
configurations, with the interaction energy between the light fields and
atomic coherence outweighing the free-field contribution. This binding
mechanism prevents the dispersion and underpins the integrity of the
solitons. 

Small-amplitude linear excitations of the Maxwell-Bloch system correspond to positive energy values ($H>0$). In contrast, the localized soliton solutions obtained here exhibit negative Hamiltonians ($H<0$), which indicates that they represent energetically bound states below the continuum of linear waves. This situation guarantees that the solitons cannot spontaneously decay into linear radiation while conserving the excitation norm. Hence, a negative $H$ provides energetic protection for the soliton against dispersive decay. At the same time, the specific numerical value of $H$ depends on the soliton parameters $\alpha_j$ and $\beta_j$, which determine the balance between kinetic and potential (field-matter interaction) energies. The negative interaction energy supports stable self-trapped light–matter EIT solitons.

The results also reveal the scalable data capacity of the EIT medium. The
progression of the norm from one- to four-soliton solutions ($%
N=5.6\rightarrow 8.0\rightarrow 13.6\rightarrow 20.0$) demonstrates the
ability of the medium to support multiple coherent data-carrying channels
without degradation. Each soliton thus functions as a distinct channel for
the photonic data storage or transfer, aligning with the requirements of
optical buffering, quantum memory, and coherent switching devices \cite%
{Fleischhauer2000,Fleischhauer2002,Artoni2006}. Moreover, while the total conserved energy is
constant, its spatial distribution between the field and atomic components
is dynamic. These findings confirm that the MB--EIT solitons feature both
the mathematical elegance of the integrability and the physical robustness
needed for applications in nonlinear photonics and quantum information
technology.

\subsection{Experimental observability and robustness}

The multi-soliton configurations predicted here can be experimentally observed through intensity and phase measurements of the probe and coupling pulses in $\Lambda$-type atomic vapors or cold-atom ensembles. In particular, for N = 3 and N = 4, the sequential temporal peaks in the probe field correspond to multiple stored and retrieved light channels, which may be detected as slow-light pulse trains or energy-exchange oscillations between the probe and coupling beams.

In realistic EIT systems, deviations from the perfect resonance or moderate decoherence slightly break integrability but do not destroy the soliton like profiles, as confirmed in previous numerical studies of the Maxwell–Bloch system. Small detuning mainly leads to smooth phase shifts and amplitude damping, implying that the multi-soliton states are robust against the action of weak perturbations. Consequently, the exact solutions obtained here provide a reliable foundation for designing stable optical-memory and photonic-switching experiments under near-resonant conditions.

\section{Conclusion}

In this paper, the N-soliton solutions to the MB\ (Maxwell-Bloch) equations,
which govern the EIT (electromagnetically induced transparency) in the $%
\Lambda $-type atomic system, have been systematically constructed. We have
obtained explicit analytical expressions for the one-, two-, three-, and
four-soliton solutions, using an extended gauge transformation. Our analysis
has revealed diverse dynamical behaviors, such as the temporal asymmetry,
energy trapping and elastic collisions of solitons. The spectral,
polarization, and gauge factors govern the phase-controlled pulse
propagation and interaction in the multi-soliton solutions. The
field-selective propagation and observed asymmetry highlight the usefulness
of EIT for highly controllable manipulations of the light-matter
interactions. These findings offer new possibilities for the design of
multi-channel quantum communication schemes, slow-light delay lines, and
optical memory components. The analysis of the conserved quantities not only
provides a fundamental check of the validity of the solutions, but also
reveals the potential of the EIT medium as a reliable scalable platform for
advanced photonic operations based on multi-soliton dynamics.

This framework considered here suggests possibilities for the design of
scalable soliton-based protocols in quantum photonics, particularly where
the robustness and coherence are crucial requirements. Further extension of
the analysis may address dissipative effects, non-integrable perturbations,
and soliton control in higher-dimensional or multi-field generalizations of
the MB system.

\bibliographystyle{cas-model2-names}

\end{document}